\documentstyle[11pt]{article} 

\newcommand{\beq}{\begin{equation}}
\newcommand{\eeq}{\end{equation}}
\newcommand{\bea}{\begin{eqnarray}}
\newcommand{\eea}{\end{eqnarray}}

\newcommand{\Ds}{\not\!\! D}

\newcommand{\aas}{/\kern-.50em A}

\newcommand{\ps}{\not\! p}

\newcommand{\ra}{\rightarrow}

\newcommand{\D}{{\cal{D}}}

\begin{document}

\author{Hugo R. Christiansen\thanks{Electronic address: 
hugo@cat.cbpf.br}\\
{\normalsize\it  Brazilian Center for Physics Research, 
CBPF - DCP}\\
{\normalsize\it Rua Xavier Sigaud 150, 22290-180\, Rio de Janeiro, 
Brazil.}}
\title
{{\bf Topological contributions to fermionic correlators and nonperturbative 
aspects of QCD in two dimensions}} 
\date{}
\maketitle

\baselineskip 18pt

\begin{abstract}
We analyze the formation of fermionic condensates in two dimensional
quantum chromodynamics for matter in
the fundamental representation of the gauge group.
We show that a topological regular instanton background is crucial
in order to obtain nontrivial correlators. We discuss both massless and 
massive cases.
\end{abstract}

\section*{Motivation}

Correlation functions are fundamental quantities in the understanding
of nonperturbative QCD and hadron physics.  
The key point is that they can be considered in terms of 
fundamental QCD fields
or in terms of physical intermediate states. For instance, 
\beq
\langle\bar\psi \psi(x)\bar\psi \psi(0)\rangle = 
\frac 1{\pi}\int ds\ \mbox{Im} K(s) D(s^{1/2},x),
\eeq
where Im $K$ is the spectral density, describing the squared matrix 
elements 
of the operators $\bar\psi\psi$ between the vacuum and all the 
hadronic 
states of mass $s^{1/2}$, and $D(s^{1/2},x)$ is the
corresponding propagator. 
The spectral density is proportional to the normalized 
cross-section of anihilation processes, hence, 
using hadronic phenomenological
data, one can extract valuable information about the microscopic 
structure of the theory. 
For example, the conservation of the quantum 
numbers of the vacuum implies that a net chiral charge is carried by
such correlators,  thus, the difference between vector
and axial correlators is entirely due to the chiral asymmetry of the 
QCD vacuum, see e.g. \cite{shur}.

One evidence of a non-perturbative vacuum structure is given by the 
Gell-Mann Oakes Renner relation (see e.g. \cite{hat})
\beq
\langle \bar u u+\bar d d\rangle =-\frac{2f_{\pi}^2 m_{\pi}^2}{(m_u+m_d)}
\eeq
since such nonzero result suggests the existence of a dynamical mass in the 
(massless) quark propagator
\beq
\langle \bar\psi\psi\rangle=-i\mbox{lim}_{y\ra x^+} 
\mbox{Tr}S_F^{full}(x,y)
\eeq
\beq
S_F^{full}(q)=\frac{A(q^2)}{\ps-M(q^2)}
\eeq 
Within perturbation theory this mass, $M(q^2)$, is zero;
therefore, condensates are particularly useful in connection with
chiral symmetry breaking coming from {\it 
non-perturbative} effects, 
for instance, due to an underlying {\it topological structure}.

The strong attractive interaction among soft quarks
and the low energy cost of creating a massless pair seem to be 
responsible
for the instability of the Fock vacuum of massless fermions, giving 
rise to
a quark anti-quark condensation. 
Therefore, in order to get some insight into the complex structure 
of the 
QCD ground-state one should consider possible mechanisms for chiral 
symmetry breaking and condensate formation. 

Quantum chromodynamics in two dimensions (QCD$_2$),
is a convenient framework to discuss this kind of phenomena since
it presents several basic features suited for a fundamental 
four-dimensional 
theory (non-abelian character,  chirality properties, etc.) and, 
furthermore, analytical results can be generally obtained.

We will use a path-integral approach which is very appropriate
to handle non-Abelian gauge theories with topological sectors
and to perform a series expansion in the fermionic mass.


\section*{\bf Decoupling and Fermionic correlators}

We work over $SU(N)$ Yang-Mills gauge fields 
coupled to massless Dirac fermions in the fundamental representation
of the gauge group, in Euclidean two-dimensional space-time:

\beq
L=\bar\psi^{q}(i\partial_{\mu} \gamma_{\mu} \delta^{qq'}+A_{\mu,a}
 t_a^{qq'}\gamma_{\mu})\psi^{q'}+
\frac{1}{4g^2} F_{\mu\nu}^a F_{\mu\nu}^a,
\label{lag}
\eeq
where the meaning of the indices is clear.
As it is done with fermions in four dimensions, we will consider a 
background of gauge fields belonging  
to nontrivial topological sectors (see next section for a more detailed
discussion).
In order to proceed to the path-integration it is worth performing
a decoupling transformation of gluons from fermions.
It is important that the decoupling operation does not change the 
topological sector in which the Dirac operator is defined. Hence,
we decompose every gauge field belonging 
to the $n^{th}$ topological sector in the form 
\beq
A_\mu^a(x) = A_\mu^a{}^{(n)} + a_\mu^a.
\label{pi}
\eeq
$A_\mu^{(n)}$ is a fixed, classical configuration of the $n^{th}$ 
topological class and $a_\mu$ is the path-integral variable which 
takes into account quantum fluctuations and belongs to the trivial 
topological sector \cite{bc}.   
Thus, the integration measure must be only defined on the $n=0$ 
sector while 
the Dirac operator depends on a regular localized topological 
background $A_\mu^{(n)}$, as we will see in what follows.

To compute fermionic correlators containing products of local bilinears 
$\bar \psi \psi(x)$, we decouple fermions from the $a_{\mu}$ field
through a chiral rotation within the topologically trivial sector, 
yielding 
a Fujikawa jacobian \cite{fuji}. The choice of an appropriate 
background like
\beq
A^{(n)}_+ = 0
\label{ba}
\eeq
is important in order to control the zero-mode problem,
so we switch to light-cone coordinates \cite{multi}.

Let us start by introducing group-valued fields to represent $A^{(n)}_{\pm}$ 
and $a_{\pm}$

\beq
A^{(n)}_- = i d \partial_- d^{-1}
\label{a}
\eeq
\beq
a_+ = i u^{-1} \partial_+ u\ \ \ \ \ \ 
a_- = i d(v \partial_- v^{-1}) d^{-1}
\label{vv}
\eeq
and accordingly define $\zeta$ by
\beq
\psi_+=dvd^{-1}\zeta_+\ \ \ \ \psi_-=u^{-1}\zeta_-
\label{lasttrafo}
\eeq
The Dirac equation then takes the form
\beq
\Ds[A^{(n)} + a]\left( \begin{array}{c}
			\psi_+ \\ 
			\psi_-
			\end{array} \right) =
 \left( \begin{array}{cc} 0 & u^{-1}i\partial_+  \\
 dvd^{-1}D_-[A^{(n)}]	 & 0 \end{array} \right)
\left( \begin{array}{c}
			\zeta_+ \\ 
			\zeta_-
			\end{array} \right)
\label{matrix2}
\eeq
Thus, the interaction Lagrangian in the $n^{th}$ flux sector 
decouples as follows
\beq
L =\bar\psi\Ds[A^{(n)}+a]\psi= \ \
\zeta_+^*\Ds_-[A^{(n)}]\zeta_+ + \zeta_-^*i\partial_+\zeta_-
\label{Lzeta}
\eeq

In terms of the representation (\ref{vv}-\ref{matrix2}) the fermionic 
determinat can be suitably factorized  by repeated use of the 
Polyakov-Wiegmann identity \cite{polw}, resulting in a gauge 
independent expression

\beq
\det \Ds[A^{(n)} + a]  = 
{\cal N}\,\det \Ds[A^{(n)}]\ \exp{-S_{eff}[u,v; A^{(n)}]}
\label{sui}
\eeq
where
\begin{eqnarray}
&& S_{eff}[u,v; A^{(n)}]  =  W[u]
+ W[v] +\frac{1}{4\pi}tr_c\int d^2x (u^{-1} 
\partial_+ u) (d \partial_- d^{-1}) \\
&&  
+\frac{1}{4\pi}tr_c\!\int\! d^2x\, (d^{-1} \partial_+ d) 
(v \partial_- v^{-1})
+\frac{1}{4\pi}tr_c\!\int\! d^2x\, 
(u^{-1} \partial_+ u)\, d\, (v \partial_- v^{-1})\, d^{-1},
\nonumber
\label{pris}
\end{eqnarray}
$W[u]$ being the usual Wess-Zumino-Witten action.
Notice that the fermionic jacobian associated with eq.(\ref{lasttrafo}) 
is precisely the quotient of the determinants  in eq.(\ref{sui}).

Once the determinant has been written in the form (\ref{sui}),
one can work with any gauge choice. The partition function shows 
the following structure
\bea
Z  &=&  \sum_n  \det(\Ds[A^{(n)}])  \int \D a_\mu\,
\Delta_{FP}\, \delta(F[a])\nonumber\\  
& & \exp \left( -S_{eff}[A^{(n)}, a_\mu] - \frac{1}{4g^2} 
\int d^2x  F^2_{\mu\nu}[A^{(n)}, a_\mu] \right)
\label{z1}
\eea
where $\Delta_{FP}\, \delta(F[a])$ comes from the gauge fixing. 

As it happens in the Abelian case, the partition function of two
dimensional quantum chromodynamics
only picks the contribution from the trivial sector because
$\det(\Ds[A^{(n)}])=0$ for $n\neq 0$ (see eq.(\ref{z1})). 
In contrast, various correlation functions become nontrivial precisely
for $n\neq 0$ thanks to the  zero-mode contributions when 
Grassman integration is performed.

In terms of the new fields, the elementary bilinear $\bar\psi\psi$
takes the following form
\beq
\bar\psi\psi\ =\ \zeta_-^*u dvd^{-1}\zeta_+ +\zeta_+^* 
(udvd^{-1})^{-1}\zeta_- 
\eeq
and after a lengthy calculation, arbitrary non-Abelian cor\-relators of 
fundamental ferm\-ions are found:
\bea
& & \langle \bar\psi\psi(x^1)\dots \bar\psi\psi(x^l)\rangle=
\sum_n\int \D u\D v\ \Delta_{FP}\ \delta (F[a_{\mu}])\, 
\exp[ -S^{Bos}_{full}(A^{(n)},u,v) ]
\nonumber\\
& & 
\int \D\bar\zeta \D\zeta\ \exp(  
\zeta_+^*\Ds_-[A^{(n)}]\zeta_+ + \zeta_-^*i\partial_+\zeta_- )
\nonumber\\
& & 
\left( B^{q_1p_1}(x^1)\dots B^{q_lp_l}(x^l)\
\zeta_-^{*q_1}\zeta_+^{p_1}(x^1)\dots\zeta_-^{*q_l}\zeta_+^{p_l}(x^l)
+B^{q_1p_1}(x^1)\dots \right.\nonumber\\
& & 
B^{-1 q_lp_l}(x^l)\ \zeta_-^{*q_1}\zeta_+^{p_1}(x^1)\dots 
\zeta_+^{*q_l}\zeta_-^{p_l}(x^l)
+B^{q_1p_1}(x^1)\dots  \nonumber\\
& & 
B^{-1 q_{l-1}p_{l-1}}(x^{l-1})B^{-1 q_lp_l}(x^l)\
\zeta_-^{*q_1}\zeta_+^{p_1}(x^1)\dots 
\zeta_+^{*q_{l-1}}\zeta_-^{p_{l-1}}(x^{l-1})
\zeta_+^{*q_l}\zeta_-^{p_l}(x^l)
\nonumber\\
& & 
\left. +\dots + 
B^{-1 q_1p_1}(x^1)\dots B^{-1 q_lp_l}(x^l)\
\zeta_+^{*q_1}\zeta_-^{p_1}(x^1)\dots\zeta_+^{*q_l}\zeta_-^{p_l}(x^l)
\right),
\label{larga}
\eea
where $B=u dvd^{-1}$.
This is a  general and completely 
decoupled expression for fermionic correlators,  which shows 
that the simple product one finds in the Abelian case becomes here 
an involved sum due to color couplings.

The fermionic path-integral can be easily performed, amounting
to a sum of products of zero modes of the Dirac operator. 
Concerning the bosonic sector, the presence of the 
Maxwell term crucially changes the effective dynamics with respect
to that of a pure Wess-Zumino model. One then has to perform 
approximate calculations  to compute the bosonic factor,
for example, by linearizing the group transformation \cite{fns};
nevertheless, the point relevant to our discussion of
obtaining nonzero ferm\-ionic correlators is manifest in eq.(\ref{larga}).

The introduction of a flavor index implies additional degrees 
of freedom which result in $N_f$ independent fermionic field variables. 
Consequently, the growing number of Grassman (numeric) differentials 
calls for additional Fourier coeficients in the integrand. 
On the other hand, dealing with $N_f$ fermions coupled to the gauge field,  
leads to the fermionic jacobian computed for one
flavor to the power $N_f$, the bosonic measure remaining the same.
For a given number of zero modes,
more fermion bilinears will be needed in order to obtain 
a nonzero fermionic path-integral. Since flavor yields a factor 
$N_f$ on the number of Grassman coeficients, the minimal nonzero
product of fermion bilinears requires a factor $N_f$ on the number
of insertions. 

\section*{Topology}
At this point, it is worth making some clarifying remarks concerning
topological gauge field configurations in this context.
In two Euclidean dimensions, finite action {\it topologically stable}
configurations of a gauge theory,
necessarily require maximal symmetry breaking,
i.e. up to $Z_N$ for the $SU(N)$ case. 
Unlike in four dimensions, where finite action solutions do exist for
pure Yang-Mills theory, true instanton {\it solutions} in 2d are 
Nielsen-Olesen type vortices \cite{no-dvs} 
(both in the Abelian and non-Abelian case) thus requiring
maximal spontaneous symmetry breaking through Higgs scalars. 
It is important to note that even with adjoint fermions, where 
the center of $SU(N)$ is immaterial (thus allowing nontrivial 
winding gluon fields in the theory) there is no topologically stable regular 
solution as it stands. 
The best one can do in that case is choosing an Abelian-like vortex which
can be taken from an effective bosonic action coming from the 
Schwinger model, as an ansatz. However, it is a meaningful procedure
just for high temperatures; at low temperatures, one has 
to construct convenient vortex configurations by hand \cite{smilga}.

For fundamental quarks, in turn, we can proceed as explained above, 
putting fermions in the background of the $Z_N$-vortices of 
de Vega and Schaposnik \cite{dvsch} which give a
realization in a two-dimensional Euclidean theory 
of regular gauge field configurations carrying a topological charge.   
This is ensured by the presence of a Higgs field in the adjoint 
which maximally breaks the gauge symmetry down to 
$SU(N)/Z_N$.

Then, if one is to consider instantons in two-dimensional 
non-Abelian gauge theories, the relevant homotopy group is 
$\pi_1(SU(N)/Z_N)$ for fermions in the adjoint as well as in the fundamental.  
The homotopy group being $Z_N$, signals the appearence of $N$ topologically 
different sectors and instanton effects become apparent. 
Gauge field configurations lying in the Cartan subalgebra of 
$SU(N)$ generate non\-trivial topological fluxes \cite{mand} 
and the index of the Dirac operator can be computed 
for fundamental fermions in a model involving the above mentioned
adjoint scalar fields  
\cite{le-gus}. 
These winding configurations may be used in the nonperturbative analysis of 
a gauge theory like QCD$_2$ when coupled to massless fermions, as we have 
shown.  
This yields important phenomenological results, especially, concerning 
chiral symmetry breaking.
Once regular gauge field configurations carrying
a topological charge are identified, the associated fermionic
zero modes can be found \cite{prd91},  
and then used to study the formation of fermion condensates.


\section*{\bf Large $N$ limit and BKT effect}

As a by-product, our approach gives $\langle \bar\psi\psi\rangle = 0$
in every flux sector, including $n=0$, as expected.  
This result is in agreement with 
independent analytical calculations based on operator product 
expansion and dispersion relations \cite{zhit}, and canonical 
quantization on the light-cone front \cite{lenz} (for any finite $N$). 
However, previous analyses have not taken into account 
topological sectors.  

In any case, despite this zero result, the existence of a  nonvanishing 
elementary condensate is desirable in the model, as it turns out in real 
QCD. Actually, such an outcome
has been put forward for one flavor QCD$_2$ within the alternative scenarios
mentioned above, 
provided one works with an infinite number of colors,
assuming that cluster decomposition holds in order to compute
this condensate from a two-point one.

In a sense, this reminds the case of massless two-dimensional QED. 
In the Abelian case, a multipoint composite
receives contributions from different topological sectors. 
In particular, one can obtain the value of the elementary scalar condensate 
from a two-point correlator. From cluster decomposition
the term $\langle \bar\psi_+\psi_+(x)\bar\psi_-\psi_-(y) \rangle$
factorizes as  $\langle \bar\psi_+\psi_+(x)\rangle \cdot
\langle\bar\psi_-\psi_-(y) \rangle$, the two factors being equal to 
each other and nonzero \cite{jaye}. Then, by means of the chiral decomposition
$\langle \bar\psi\psi(x)\rangle =\langle \bar\psi_+\psi_+(x)
\rangle+ \langle\bar\psi_-\psi_-(x) \rangle$ one is able to construct the  
condensate from the trivial topological sector whereas,  in fact, 
$\langle\bar\psi_{\pm}\psi_{\pm}(x)\rangle$ come exclusively from 
flux classes $\pm 1$  respectively.
Now, we would like to find analogous features in two-dimensional QCD,
in order to emphasize the role played by instanton contributions in the 
non-Abelian counterpart, as it seems to occur in four dimensions. 
We will see below that this is possible.

In the non-Abelian theory, 
one needs the large $N$ limit for cluster decomposition to take place, 
so that all v.e.v. can be reduced to a product of elementary scalar 
densities,
i.e. $\langle A B \rangle = \langle A \rangle\langle B \rangle + O(1/N)$.
On the other hand, for very separated composites factorization also holds.

By means of QCD sum rules, in the weak coupling regime
it can be shown that a BKT effect (Berezinskii-Kosterlitz-Thoules \cite{bkt}) 
takes place \cite{zhit}.  For very large distances and number of colors:
 
\beq
\langle \bar\psi_+\psi_+(x)\bar\psi_-\psi_-(y)\rangle\sim |x-y|^{-1/N}
\label{bkt}
\eeq
implying that for large, but finite $N$, it smears away softly as  
$|x-y|\ra \infty$. Though, when $N$ is infinity, it is nonzero.

In fact, this effect has been found 
when the $M \ra 0$ limit is taken at the end of the calculation
in the {\em massive} theory, provided $M>>g\sim 1/\sqrt{N}\ra 0$
('t Hooft regime).

One could then say that after the chiral weak phase is reached
in the massive theory,
the symmetry still keeps in a broken phase, but instead, 
it happens dynamically.
In the large $N$ limit the physics changes,  the vacuum  being 
dressed non-perturbatively by means of purely planar diagrams. 
The spectrum is completely different depending on the relations 
shown above:
In the weak phase there is an infinite number of massive mesons
while in the strong coupling phase there are just massless baryons.

Outside any of the asymptotic relations above,
the actual results, for arbitrary values of both color and 
relative positions, are given by eq.(\ref{larga}),
provided topological sectors are included.
As already announced by the chiral jacobian given by (\ref{sui}), the 
existence of nontrivial correlators  signals that a 
chiral anomaly takes place in the model. This is expected according
to the Coleman's theorem, which prohibits the spontaneous breakdown 
of continuous symmetries in two dimensions \cite{col-th}.


\section*{The massive case}

Now, in order to find out the above nonzero outcome for the elementary 
condensate in our approach, we will extend our procedure to massive 
QCD$_2$ as follows. The partition function now is
\beq
Z_M = \int \D \bar\psi \D \psi\D A_\mu \exp(-\int d^2x\,  L_{M=0})
\ e^{-M \int d^2x\, \bar\psi\psi} 
\label{zm}
\eeq
so that, it is apparent that a series expansion may be workable \cite{mass}.
Actually, the analytical solution to this model has not been found
and such a challenge  might be a hopeless effort.
Nevertheless, for our purposes only small masses need to be examined
so we may perform a perturbative expansion in terms of the quark mass.  
Then, the minimal condensate now reads

\bea
& & \langle\bar\psi\psi(\omega)\rangle_M =\sum_n
\langle\bar\psi\psi(\omega)\rangle_{M=0}^{(n)} + 
M\ \sum_n \int d^2x\, \langle\bar\psi\psi(\omega)\bar\psi\psi(x)
\rangle_{M=0}^{(n)} +
\nonumber\\
& & 
\frac{1}{2} M^2\ \sum_n \int d^2x\, d^2y
\, \langle\bar\psi\psi(\omega)\bar\psi\psi(x)
\bar\psi\psi(y)\rangle_{M=0}^{(n)} + \dots
\label{cm2}
\eea
In this fashion, it is apparent that for $M\neq 0$ this elementary 
condensate
receives contributions from every correlator coming from the massless
theory.  As we have seen in the first
section, this is a pleasant result in order to mimic real four-dimensional 
strong interactions among quarks, where an instanton vacuum seems to 
be responsible for the chiral symmetry breaking.

Since the fermionic sector has been completely decoupled, the counting 
of the nonzero pieces simply follows from that of the Abelian case 
\cite{steele,hf}
because the topological structure here can be red out from the 
torus of the gauge group.

In the compactified space, the existence of $n N$ normalizable zero 
modes in topological sector $n$ \cite{prd91,le-gus} implies
the vanishing of the first summatory in eq.(\ref{cm2}) $\forall n$. 
However, for higher powers of $M$ it is clear that certain nonzero
contributions come into play. 
As we have explained, the existence of zero modes with
a definite chirality (positive for $n>0$, negative for $n<0$) set
the (Grassman) integration rules to compute v.e.v's.
 
By using the chiral decomposition 
$\langle \bar\psi\psi(x)\rangle =\langle \bar\psi_+\psi_+(x)
\rangle+ \langle\bar\psi_-\psi_-(x) \rangle$
one can see that the first $N$ powers of $M$
($j= 1\dots N$) receive an input from the trivial 
topological sector alone.
Then, for $j\geq N$, the $n=1$ sector starts contributing 
together with $n=0$.
For powers $j\geq 2N$  the contribution of topological 
sector $n=2$ starts, and the counting follows so on in this way.

Now, it can be easily seen that the number of contributions grows 
together with the number of colors. 
As we let $N$ go to infinity the elementary `massive' condensate does so.
On the other hand, since within each nontrivial topological sector
the number of zero modes grows also to infinity, one has
divergent v.e.v. everywhere in the series expansion of eq.(\ref{cm2}).
Accordingly, high order terms could also produce a nonzero outcome
in the chiral limit.
Therefore, it is clear that the limit $M\ra 0$ becomes matter of a
careful analysis; namely, combined with a large number
of colors,  eq.(\ref{cm2}) leaves place 
enough for a nontrivial elementary condensate even in the chiral limit. 


\section*{\bf Summary}

General fermionic condensates have been computed
in non-Abelian gauge theories  
in order to discuss chiral symmetry breaking and vacuum properties of QCD
in two dimensions. 

We have shown that correlation functions can be completely determined 
by the topological structure underlying the theory.
As we have explained, regular gauge fields lying in the Cartan 
subalgebra of $SU(N)$ have to be taken into
account to find a significant outcome for general fermionic correlators.

As it was discussed at the beginning of this paper, this is a welcome 
result,
inasmuch our scheme enables a chiral symmetry breaking mechanism
at the vacuum level,  by means of instanton contributions.
This is clearly reflected in the quantum calculation, which exhibits 
the formation of
nonvanishing vacuum expectation values of fermionic bilinears.

Finally, our analysis of the massive theory put forward 
possible roots to the BKT phenomenon coming from 
topological considerations, by means of a path-integral approach. 
In this way, although an exact result has not been given so far, both the 
zero outcome in the massless theory and the nontrivial value in the massive
one, have gained contact with the results emerging from alternative 
scenarios \cite{zhit,lenz,mi,zhit2}.

\section*{Acknowledgments}  I would like to thank T. Hatsuda, F. Lenz, 
F. Schaposnik, E. Shuryak and I. Zahed for enlightening comments
and suggestions. This work was supported by CBPF and CLAF-CNPq, Brazil.


\end{document}